\documentclass[11pt]{elsarticle}
\makeatletter
\def\ps@pprintTitle{%
	\let\@oddhead\@empty
	\let\@evenhead\@empty
	\def\@oddfoot{}%
	\let\@evenfoot\@oddfoot}

\usepackage{amssymb}
\usepackage{amsfonts}
\usepackage{amsmath}
\usepackage{latexsym,epsfig}
\usepackage{graphicx}
\usepackage[usenames,dvipsnames]{color}
\usepackage{url} 
\usepackage{soul} 
\usepackage{array}
\usepackage{verbatim}
\usepackage{hyperref}

\DeclareMathOperator\Exp{Exp}

\usepackage[numbers]{natbib}
\begin{document}

\begin{frontmatter}

\title{Measuring privacy in smart metering anonymized data}

\author[udl]{Santi Mart\'{\i}nez}
\ead{santi@matematica.udl.cat}

\author[udl,cybercat]{Francesc Seb\'e}
\ead{fsebe@matematica.udl.cat}

\author[uds]{Christoph Sorge}
\ead{christoph.sorge@uni-saarland.de}

\address[udl]{Department of Mathematics. Universitat de Lleida. C. Jaume II, 69, E-25001 Lleida (Spain).}

\address[cybercat]{Center for Cybersecurity Research of Catalonia.}

\address[uds]{Saarland University, P.O. Box 15 11 50, D-66041 Saarbrücken (Germany).}

\begin{abstract}
In recent years, many proposals have arisen from research on privacy in smart metering. One of the considered approaches is anonymization: smart meters transmit fine-grained electricity consumption values in such a way that the energy supplier can not exactly determine procedence. This paper measures the real privacy provided by such approach by taking into account that at the end of a billing period the energy supplier collects the overall electricity consumption of each meter for billing purposes. An entropy-based measure is proposed for quantifying privacy and determine the extent to which knowledge on the overall consumption of meters allows to re-identify anonymous fine-grained consumption values.
\end{abstract}


\end{frontmatter}

\section{Introduction}
\label{sec:intro}

Smart meters measure and transmit information about electricity consumption to the energy supplier in short intervals (each 15 or 30 minutes). Such data allows an accurate prediction of consumption so that the production can be managed in real-time. The replacement of old analog meters with smart ones is in progress in many developed countries. For instance, the European Union expects to replace 80\% of electricity meters by 2020. In Spain, such deployment was almost complete by the end of 2018.

Unfortunately, such fine-grained data about household electricity consumption allows to infer sensitive information like the types of electrical devices being used, or even the exact times people leave or arrive at home.

Encryption of electricity readings prior to their transmission protects against external eavesdroppers but not against a malicious energy supplier which may process them for non-authorized purposes. Hence, privacy-preserving solutions  for smart metering data transmission should assume that the energy supplier is an untrusted entity.

Solutions providing privacy in smart metering can be classified into the following main classes:

\begin{itemize}

\item {\em Obfuscation-based methods:} Controllable batteries and alternative generation devices~\cite{KEDLC10}, or devices with an adjustable consumption~\cite{egarter2014load} are placed inside the household so as to alter energy usage curves.

\item {\em Perturbative methods:} Meters mask a fine-grained electricity consumption value by adding some random noise to it prior to its transmission to the energy supplier.
In this way, the energy supplier gets only an approximation to the real household consumption profiles.
The type and magnitude of the added noise has to be accurately tuned to enforce privacy while preserving the
accuracy of the transmitted data.  The proposal~\cite{BBA2016} provides differential privacy~\cite{D08} by LaPlacian noise addition. The proposal~\cite{BSU2010} is based on adding Gaussian noise.

\item {\em Anonymous methods:} Data are transmitted so that the link between electricity readings and the identity of customers is removed. The proposals~\cite{EK10,FB13} propose the use of pseudonyms for sending high frequency metering data. In the aforementioned proposals each household has a unique pseudonym which is transmitted together with all its consumption values so that the electricity providers are able to get anonymous consumption patterns (linked to pseudonyms). In~\cite{JJR11} it is proven that pseudonymized consumption traces can be re-identified if combined with external indicators. The authors show the high re-identification performance of two attack vectors called {\em anomaly detection} and {\em behavior pattern matching}. Both attacks are shown to be robust against basic countermeasures like resolution reduction or frequent re-pseudonymization.

A different approach is taken in~\cite{P10,SK2012}. In these proposals, the electricity consumptions received by the electricity supplier at a given period come without a pseudonym~\cite{P10} or linked to a pseudonym shared among several meters~\cite{SK2012}. In this way, a given consumption value could come from any meter belonging to the originating community.

\item {\em Aggregation-based methods:} 
The meters are grouped into neighborhoods or communities with a substation which has direct communication with each meter. Each metering period, the meters and the substation engage in a protocol which results in the substation receiving the aggregated (added) readings of all the meters.
Such proposals employ homomorphic cryptosystems for ensuring the substation gets no information but the addition of all meters readings in a metering period.
This problem, when not focused on smart metering data,  is referred to as {\em privacy-preserving aggregation of time-series data}~\cite{JL13,RN10,SCCSR11}.

The authors in~\cite{GJ2011} consider a neighborhood with $n$ smart meters, $sm_1,\ldots,sm_n$ and a substation. Each meter stores a private key whose related public key is publicly known.
Each meter takes its energy consumption $e_{i}$ at a given period and generates $n$ values $m_{i,j}$, $j \in \{1,\ldots,n\}$,
satisfying $e_{i}=m_{i,1} + \ldots + m_{i,n}$. Then each meter $sm_i$ encrypts each value $m_{i,j}$, $j\neq i$ under the public key of $sm_j$ and
transmits the resulting ciphertext to the substation. The substation homomorphically aggregates all $n-1$ ciphertexts encrypted under the
public key of each meter and sends the result to it. Each meter $sm_{i}$, decrypts the received ciphertext and adds $m_{ii}$ to it. Finally,
it sends the result to the substation. The substation gets the aggregated consumption as the addition of the results received from all the meters.  The proposal has a high, $O(n^2)$, communication cost per execution.

The authors in~\cite{RN10} suggest a method which can be applied to aggregate smart meter readings.
There exists a neighborhood public key whose private key is distributed among all the meters.
Each meter in a neighborhood encrypts its consumption value using the Paillier cryptosystem under the neighborhood public key. The substation aggregates all the encrypted consumptions and then requests
each meter to compute a partial decryption of the resulting ciphertext. From all the partial decryptions, the
substation obtains the aggregated consumption. The proposal includes a method to prevent a malicious
coalition including the substation and some corrupted meters from getting information about the consumption of honest meters other than their aggregated sum.
The authors also include a way to add LaPlacian noise to the consumptions so as to get differential privacy.
A drawback of this proposal comes from the use of a distributed Paillier cryptosystem whose setup is very complicated when a trusted dealer is not desirable~\cite{NS11}.
The proposal~\cite{BPSSV16}, enhanced in~\cite{Garra2019}, proposes a similar aggregation mechanism which makes use of
the homomorphic property of ElGamal cryptosystem with a quite easier distributed private key setup.

The proposals~\cite{SCCSR11,XZ15} require the presence of a trusted dealer that generates a set of random
values that sum up to zero. As pointed out in~\cite{JungLi15}, and implemented in~\cite{GMMS18}, presence of a trusted dealer during setup can be avoided by making use of a trusted dealer--free protocol for secure addition like~\cite{GJ2011,RN10,LLL10}. Each meter and the substation receive one of such values. Each meter uses its secret share to encrypt its reading prior to
transmitting it. The substation aggregates the received encrypted data and gets the
addition of all the readings after solving an easy instance of the discrete logarithm problem.
The authors in~\cite{SCCSR11} extend their basic proposal to provide differential privacy by adding symmetric geometric noise.
The proposal is very lightweight and only requires unidirectional meter to substation data transmission.
Unfortunately, in dynamic scenarios, each time a meter is added or removed, the presence of a trusted dealer or execution of a trusted dealer--free protocol for secure addition is required. 
A similar proposal which avoids the computation of a discrete logarithm after decryption at the cost of requiring a trusted dealer that generates an RSA modulus $N=pq$, is proposed in~\cite{JL13}. A proposal equivalent to~\cite{JL13} can be found in~\cite{BM14}.

\end{itemize}

The Germany's information security agency has designed a smart metering system architecture which considers the privacy of consumers. The solution belongs to the {\em anonymous methods} paradigm.
As discussed in~\cite{SK2012}, the key component of that architecture is a smart metering gateway which is installed in a consumer's house. The main tasks of a gateway are to get local meters readings and communicate remotely with the energy service provider and the grid operator. Remote communications can be identifiable or pseudonymised.

Identifiable communications are used for transmitting the sum of electricity consumption values (e.g., once per month) to the service provider for billing purposes. Pseudonymised communications are for privately sending fine-grained electricity consumption values. First, the gateway generates a message composed of the consumption value together with a pseudonym. That message is encrypted under the grid operator's public key. Next, the resulting ciphertext is digitally signed and transmitted to a gateway operator which verifies the digital signature, removes it, and forwards the encrypted message to the grid operator. The grid operator finally decrypts the received message and gets the consumption value together with the pseudonym. So as to keep the source of data secret, a given pseudonym must be shared among a certain amount of gateways. A pseudonym shared among $k$ gateways is assumed to provide $k$-anonymity since a dishonest grid operator can not determine which of the $k$ gateways sharing a pseudonym is the source of a given consumption value.

\section{Problem statement}
\label{sec:statement}

As detailed in the model~\cite{BSU2010}, a smart metering application consists of an electricity supplier (ES) and a set of smart meters $S= \{sm_1,\ldots,sm_n \}$. Let $e_{i,j}$ be the electricity consumption measured by $sm_i$ in period $j$. We denote by $t$ the number of time periods included in a billing period.
In a scenario in which consumption values are sent sharing the same pseudonym, at the end of a given billing period, the information collected by the ES is:
\begin{itemize}
 \item For each meter $sm_i$, the sum of its electricity consumption values over the billing period.
That is, $E_i = \sum_{j=1}^t e_{i,j}.$

\item For each period $j$, the set of pseudonymized electricity consumption values transmitted by all the meters in $S$.
We denote these values as $\{e'_{1,j},\ldots, e'_{n,j} \}$.
Note that there exists a permutation $\pi_j$, unknown to the ES, so that $e_{i,j} = e'_{\pi_j(i),j}$.
\end{itemize}

The privacy of customers is preserved as long as each permutation $\pi_j$ stays secret. Hence, the objective of
a malicious ES aiming to get information about the consumption habits of a given meter $sm_1$ is to determine $\pi_j(1)$ for each time period $j \in \{1,\ldots, t\}$, so that the consumption values of $sm_1$ can be reidentified as $\{e_{1,1},\ldots,e_{1,t} \} = \{e'_{\pi_1(1),1},\ldots,e'_{\pi_t(1),t} \}$.

\subsection{Example}

Let us assume a smart metering application composed of three smart meters $\{sm_1, sm_2, sm_3\}$ and a billing period composed of nine time periods. The aggregated consumptions of the meters are $E_1 = 991$, $E_2=473$ and $E_3=926$.
The pseudonymized consumption values are shown in Table~\ref{tab:table1}.

\begin{table}[h!]
  \centering
  \begin{tabular}{c|l}
    Period & Consumptions\\
    \hline
    1 & $\{117, 104, 362 \}$ \\
    2 & $\{89, 50, 64 \}$ \\
    3 & $\{25,  119, 86 \}$ \\
    4 & $\{23, 25, 149 \}$ \\
    5 & $\{86, 140, 49 \}$ \\
    6 & $\{36, 87, 117  \}$ \\
    7 & $\{42, 146, 108 \}$ \\
    8 & $\{24, 83, 92 \}$ \\
    9 & $\{56, 24, 87 \}$ \\
  \end{tabular}
\caption{Pseudonymized consumptions (in Wh).}
\label{tab:table1}
\end{table}

Now the ES can search for sets of three-element permutations $\{\pi_1,\ldots, \pi_9\}$ satisfying that
$\sum_{j=1}^{9} e'_{\pi_j(1),j} = 991$,  $\sum_{j=1}^{9} e'_{\pi_j(2),j} = 473$, and $\sum_{j=1}^{9} e'_{\pi_j(3),j} = 926$. There exist three such solutions shown in Table~\ref{tab:solutions_original}.

\begin{table}[h!]
  \centering
\begin{tabular}{|rcl|}
\hline
$362 + 64 + 119 + 23 + 140 + 36 + 108 + 83 + 56$ &=& $991,$ \\
$117 + 50 + 25 + 25 + 49 + 117 + 42 + 24 + 24$ &=& $473,$ \\
$104 + 89 + 86 + 149 + 86 + 87 + 146 + 92 + 87$ &=& $926.$\\
\hline
$362 + 64 + 86 + 25 + 140 + 36 + 108 + 83 + 87$ &=& $991,$\\
$117 + 50 + 25 + 23 + 49 + 87 + 42 + 24 + 56$ &=& $473,$\\
$104 + 89 + 119 + 149 + 86 + 117 + 146 + 92 + 24$ &=& $926.$\\
\hline
$362 + 89 + 86 + 25 + 140 + 36 + 146 + 83 + 24$ &=& $991,$ \\
$117 + 50 + 25 + 23 + 49 + 87 + 42 + 24 + 56$ &=& $473,$ \\
$104 + 64 + 119 + 149 + 86 + 117 + 108 + 92 + 87$ &=& $926.$ \\
\hline
\end{tabular}
\caption{Solutions to the problem.}
\label{tab:solutions_original}
\end{table}

All the three possible solutions satisfy that $\pi_1(1) = 3$, hence the ES can deduce that $e_{1,1}=362$.  In the same way, the ES also deduces that $e_{1,5}=140$, $e_{1,6}=36$, and $e_{1,8}=83$, so that the consumption of $sm_1$ at four periods is revealed.
Regarding $sm_2$, the ES can exactly determine six consumption values, namely 
$e_{2,1}=117$, $e_{2,2}=50$, $e_{2,3}=25$, $e_{2,5}=49$, $e_{2,7}=42$, and $e_{2,8}=24$.
For $sm_3$, the ES can infer $e_{3,1}=104$, $e_{3,4}=149$, $e_{3,5}=86$, and $e_{3,8}=92$.
Hence, partial information about the consumption habits of the three smart meters actually leaked.

This example shows that a smart metering application composed of $k$ smart meters sharing the same pseudonym 
can not guarantee to achieve $k$-anonymity.

\subsection{Relaxed problem statement}

Let a smart metering application with $n$ meters with a billing period composed of $t$ periods. For each period $j$ there exist $n!$ possible $n$-element permutations. Hence the search space for the previous problem is $(n!)^t$. This makes the problem solvable through exhaustive search only for instances with very small values of $n$ and $t$. In our experiments, even a dynamic programming optimized implementation of the problem has failed to provide reasonable running times.

We consider a relaxed version of the problem in which the ES focuses on just one of the meters, for instance $sm_1$.
Now, for each time period $j$, the ES will reduce its search to sets of integers $\{\pi_1(1),\ldots, \pi_9(1)\}$ satisfying
that $\sum_{j=1}^{9} e'_{\pi_j(1),j} = E_1$.

In this case, for each period $j$, there exist $n$ possible values for $\pi_j(1)$ and the search space becomes $n^t$.
Although the search space keeps being exponential, the problem aiming to find one of such solutions corresponds to an instance of the multiple-choice subset-sum problem (MCSSP) which, as stated in~\cite{P99}, can be solved in polynomial time
when the weights and profits (in our case the consumption values) are bounded by a constant.
In our case, we need to find all such solutions, but dynamic programming techniques allow us to efficiently solve even moderately large instances of the problem.

\subsection{Measuring privacy}

As said before, in a smart metering application with $n$ meters and $t$ time periods, the objective of an attacker aiming to compromise the privacy of $sm_1$ is to determine $\pi_j(1)$ for each period $j \in \{1,\ldots,t\}$. From the attacker point of view, each $\pi_j(1)$ is a random variable whose sample space is $\{1,\ldots,n\}$.

When the attacker is able to exactly determine $\pi_j(1)$, then the entropy (uncertainty) on $\pi_j(1)$ is 0 bits. On the other side, when it has no information at all about it, since there are $n$ possible values for $\pi_j(1)$, then its entropy is $\log_2 n$ bits. An attacker with partial information about plausible values for $\pi_j(1)$ will get an entropy ranging between $0$ and $\log_2 n$.

\subsection{Our approach}

In our approach, we search for all the solutions $\{\pi_1(1),\ldots, \pi_9(1)\}$ satisfying $\sum_{j=1}^{9} e'_{\pi_j(1),j} = E_1$. Then, for each $\pi_j$, we count the amount of times that each value from its range ($1,\ldots,n$) appears in a solution. Finally, we assign a probability to each solution which is proportional to its appearance rate, and compute the entropy of random variable $\pi_j$.

\subsection{Example}

In our example, in its reduced version focused on meter $sm_1$,
an attacker finds twenty-two solutions to the problem, which are shown in Table~\ref{tab:solutions_relaxed}.

\begin{table}[h!]
  \centering
\begin{tabular}{|rcl|}
\hline
$362 + 64 + 119 + 23 + 140 + 36 + 108 + 83 + 56$ & = & $991$ \\
$117 + 64 + 119 + 149 + 140 + 117 + 146 + 83 + 56$ & = & $991$ \\
$362 + 64 + 119 + 25 + 49 + 87 + 146 + 83 + 56$ & = & $991$ \\
$362 + 64 + 25 + 149 + 86 + 117 + 108 + 24 + 56$ & = & $991$  \\
$362 + 50 + 119 + 149 + 49 + 36 + 146 + 24 + 56$ & = & $991$  \\
$362 + 89 + 86 + 25 + 86 + 117 + 146 + 24 + 56$ & = & $991$ \\
$362 + 89 + 86 + 25 + 86 + 87 + 108 + 92 + 56$ & = & $991$ \\
$362 + 89 + 25 + 149 + 140 + 36 + 42 + 92 + 56$ & = & $991$ \\
$362 + 50 + 86 + 23 + 140 + 36 + 146 + 92 + 56$ & = & $991$ \\
$362 + 64 + 25 + 149 + 140 + 36 + 108 + 83 + 24$ & = & $991$  \\
$362 + 89 + 86 + 25 + 140 + 36 + 146 + 83 + 24$ & = & $991$ \\
$362 + 64 + 86 + 23 + 86 + 117 + 146 + 83 + 24$ & = & $991$ \\
$362 + 64 + 119 + 25 + 140 + 87 + 146 + 24 + 24$ & = & $991$ \\
$362 + 64 + 86 + 149 + 49 + 87 + 146 + 24 + 24$ & = & $991$ \\
$362 + 89 + 119 + 23 + 49 + 87 + 146 + 92 + 24$ & = & $991$ \\
$362 + 50 + 119 + 25 + 86 + 87 + 146 + 92 + 24$ & = & $991$ \\
$362 + 64 + 86 + 25 + 140 + 36 + 108 + 83 + 87$ & = & $991$ \\
$362 + 64 + 119 + 149 + 49 + 36 + 42 + 83 + 87$ & = & $991$ \\
$362 + 89 + 25 + 23 + 140 + 36 + 146 + 83 + 87$ & = & $991$ \\
$362 + 64 + 119 + 23 + 49 + 117 + 146 + 24 + 87$ & = & $991$  \\
$362 + 89 + 25 + 25 + 86 + 117 + 108 + 92 + 87$ & = & $991$ \\
$362 + 64 + 119 + 23 + 49 + 87 + 108 + 92 + 87$ & = & $991$ \\
\hline
\end{tabular}
\caption{Solutions to the relaxed problem.}
\label{tab:solutions_relaxed}
\end{table}

From these results, we estimate that the probability $P[\pi_1(1)=3]$ (which implies $e_{1,1}=362$) is $21/22$, while $P[\pi_1(1)=1]$ (with $e_{1,1}=117$) is $1/22$. In this case, the entropy on $\pi_1(1)$ is as low as $0.2668$ bits. 

A higher entropy is obtained for $\pi_4(1)$. In that case,
$P[\pi_4(1) = 1]$ is $7/22$,  $P[\pi_4(1) = 2]$ is $8/22$, and $P[\pi_4(1) = 3]$ is $7/22$.
In this case, the entropy we get on $\pi_4(1)$ is maximum, namely $1.582$ bits.

\section{Experimental results}

This section shows the results obtained from several experiments over real and synthetic data.

\subsection{Experiments over real data}

In our experiments we have used data from the set ``Electricity Smart Meter CBT'' kindly provided by the Irish Social Science Data Archive (ISSDA)\footnote{{\tt https://www.ucd.ie/issda/}}. That data set contains data collected from several houselholds in $30$ minute intervals. Electricity readings are provided in kWh with three decimal precision, which is equivalent to providing them in Wh.

In order to evaluate the privacy obtained for different group sizes and amount of time periods we have first filtered the data set.
The provided file contains the readings of $1000$ smart meters during about $25000$ time periods. For each experiment, we have selected a random subset of $n$ smart meters $S = \{sm_1, \ldots, sm_n\}$, and $t$ consecutive time periods.

We generated several problem instances with $n \in \{2,4,8,16\}$ and $t \in \{15, 30, 60\}$.
For each instance, we ran our solver so as to find all the solutions focusing on the first smart meter $sm_1$.
Then, we computed the entropy at each time period.

\begin{table}[htb]
 \centering
 \begin{tabular}{|c||c|c|c|c|}
  \hline
             & $n= 2$ & $n= 4$ & $n= 8$ & $n=16$ \\ \hline \hline
Max. entropy & $1.00$ & $2.00$ & $3.00$ & $4.00$ \\ \hline \hline
  $ t = 15 $ & $0.20$ & $1.93$ & $2.97$ & $3.33$ \\ \hline
  $ t = 30 $ & $0.99$ & $1.71$ & $2.68$ & $4.00$ \\ \hline
  $ t = 60 $ & $0.85$ & $1.42$ & $2.75$ & $3.64$ \\ \hline
 \end{tabular}
 \vspace{1mm}
 \caption{Average entropy for experiments over real data.}
 \label{tab:RealData}
\end{table}

Table~\ref{tab:RealData} summarizes the results obtained from these experiments.
Each column corresponds to a value for $n$.
The first row shows the maximum obtainable entropy for each $n$, namely $\log_2 n$ bits.
The rest of the rows correspond to different values for $t$. 

As it can be seen, a maximum (or almost maximum) entropy is reached in some problem instances, but, in general, the result is about $0.5$ bits below. 

Regarding the influence of the amount of time periods, $t$, it seems that this parameter has little influence on the results.
In some cases, the highest entropy has been obtained for $t=15$, while others required $t=30$. Notice that for $n=8$, $t=60$ provides a result better than $t=30$ (although in that case $t=15$ was the best).

Other problem instances were analyzed with similar results. Due to the nature of this data set, it is difficult to extract precise conclusions.

Solving instances with very large values for $n$ and $t$ is very difficult due to the high requirements both in time and space. Hence it may be possible that the adversary is even unable to obtain a set of possible solutions.

\subsection{Experiments over synthetic data}

We next detail several experiments conducted over synthetic data. So as to generate synthetic data sets we 
first inferred the probability distribution of the smart meter readings.

We analyzed the data from a single meter so as to estimate the parameters for several possible distributions.
When possible, we used the uniformly minimum-variance unbiased estimators (UMVUE) \cite{VoinovN-1993-unbias,LamSW-1994-aism}.
Then, we used Cram{\'e}r-von Mises criterion~\cite{Cramer-1928-saj,vonMises-1928-wsw} to evaluate the goodness of fit of each of the considered distributions.
We repeated this process with data from other smart meters to verify the consistency of results.

Our conclusion was that, in general, smart meter readings follow an exponential distribution (although for some meters the Cram{\'e}r-von Mises criterion pointed more to a normal distribution).

Thus, we created our synthetic problem instances as follows:
The first smart meter (the one we are trying to identify) follows an exponential distribution $\Exp(\lambda)$ with $\lambda^{-1}$ being the mean of the distribution.
Each of the remaining $n-1$ meters follows an exponential distribution $\Exp(0.01)$, so that their readings have a mean value of $100$. Several problem instances were generated at random under the mentioned distributions.

\begin{table}[htb]
 \centering
 \begin{tabular}{|c||c|c|c|c|c|}
  \hline
             & $n= 2$ & $n= 4$ & $n= 8$ & $n=16$ & $n=32$ \\ \hline \hline
  $ t = 15 $ & $0.00$ & $0.73$ & $1.73$ & $2.63$ & $3.74$ \\ \hline
  $ t = 30 $ & $0.00$ & $1.31$ & $1.75$ & $2.85$ & $4.10$ \\ \hline
  $ t = 60 $ & $0.43$ & $1.02$ & $1.99$ & $2.97$ & $3.87$ \\ \hline
 \end{tabular}
 \vspace{1mm}
 \caption{Average entropy for experiments over synthetic data for $\lambda^{-1}=20$}
 \label{tab:SyntL20}
\end{table}

\begin{table}[htb]
 \centering
 \begin{tabular}{|c||c|c|c|c|c|}
  \hline
             & $n= 2$ & $n= 4$ & $n= 8$ & $n=16$ & $n=32$ \\ \hline \hline
  $ t = 15 $ & $0.00$ & $1.54$ & $2.76$ & $3.71$ & $4.66$ \\ \hline
  $ t = 30 $ & $0.62$ & $1.47$ & $2.78$ & $3.73$ & $4.81$ \\ \hline
  $ t = 60 $ & $0.73$ & $1.80$ & $2.85$ & $3.77$ & $4.66$ \\ \hline
 \end{tabular}
 \vspace{1mm}
 \caption{Average entropy for  experiments over synthetic data for $\lambda^{-1}=50$}
 \label{tab:SyntL50}
\end{table}

\begin{table}[htb]
 \centering
 \begin{tabular}{|c||c|c|c|c|c|}
  \hline
             & $n= 2$ & $n= 4$ & $n= 8$ & $n=16$ & $n=32$ \\ \hline \hline
  $ t = 15 $ & $0.97$ & $1.99$ & $3.00$ & $3.99$ & $4.96$ \\ \hline
  $ t = 30 $ & $1.00$ & $1.98$ & $2.99$ & $3.98$ & $4.96$ \\ \hline
  $ t = 60 $ & $1.00$ & $2.00$ & $3.00$ & $4.00$ & $4.99$ \\ \hline
 \end{tabular}
 \vspace{1mm}
 \caption{Average entropy for  experiments over synthetic data for $\lambda^{-1}=100$}
 \label{tab:SyntL100}
\end{table}

\begin{table}[htb]
 \centering
 \begin{tabular}{|c||c|c|c|c|c|}
  \hline
             & $n= 2$ & $n= 4$ & $n= 8$ & $n=16$ & $n=32$ \\ \hline \hline
  $ t = 15 $ & $0.78$ & $1.61$ & $2.63$ & $3.86$ & $4.27$ \\ \hline
  $ t = 30 $ & $0.85$ & $1.78$ & $2.57$ & $3.75$ & $4.44$ \\ \hline
  $ t = 60 $ & $0.73$ & $1.78$ & $2.81$ & $3.51$ & $4.43$ \\ \hline
 \end{tabular}
 \vspace{1mm}
 \caption{Average entropy for  experiments over synthetic data for $\lambda^{-1}=200$}
 \label{tab:SyntL200}
\end{table}

\begin{table}[htb]
 \centering
 \begin{tabular}{|c||c|c|c|c|c|}
  \hline
             & $n= 2$ & $n= 4$ & $n= 8$ & $n=16$ & $n=32$ \\ \hline \hline
  $ t = 15 $ & $0.00$ & $1.31$ & $2.20$ & $3.03$ & $3.25$ \\ \hline
  $ t = 30 $ & $0.52$ & $1.22$ & $1.56$ & $2.26$ & $3.00$ \\ \hline
  $ t = 60 $ & $0.52$ & $0.82$ & $1.42$ & $2.47$ & $2.88$ \\ \hline
 \end{tabular}
 \vspace{1mm}
 \caption{Average entropy for  experiments over synthetic data for $\lambda^{-1}=500$}
 \label{tab:SyntL500}
\end{table}

Tables~\ref{tab:SyntL20}, \ref{tab:SyntL50}, \ref{tab:SyntL100}, \ref{tab:SyntL200}, \ref{tab:SyntL500} show the results obtained from these new experiments.

In each table, the data of the first smart meter $sm_1$ was generated under a different value for $\lambda^{-1}$.
Parameter $\lambda^{-1}$ was $20$ in Table~\ref{tab:SyntL20}, $50$ in Table~\ref{tab:SyntL50}, $100$ in Table~\ref{tab:SyntL100}, $200$ in Table~\ref{tab:SyntL200}, and $500$ in Table~\ref{tab:SyntL500}.

The results were similar to those obtained from real data from the set provided by the ISSDA.
As before, except for the extreme case with only $n=2$ smart meters, the amount of periods $t$ did not affect very much.
It seems, however, that $t=60$ has a larger tendency to produce the highest entropy, but that was not the case for larger values of $\lambda^{-1}$. This is specially noticeable in Table~\ref{tab:SyntL500}.

The experiments show that the attainable entropy depends on the value of $\lambda^{-1}$. It is higher when the distribution of the targeted meter is close to that of the remaining meters, and decreases as its distribution differs from the others.

When $\lambda^{-1}=100$ (Table~\ref{tab:SyntL100}), which corresponds to the case in which all the meters follow the same distribution, the highest privacy is obtained, as a maximum or almost maximum entropy for all combinations of $n$ and $t$.

When the mean of the targeted meter readings is half of (Table~\ref{tab:SyntL50}, $\lambda^{-1}=50$) or doubles (Table~\ref{tab:SyntL200}, $\lambda^{-1}=200$) the mean of the others, the highest attainable entropy is about $0.2$ bits below the maximum.

For $\lambda^{-1}=20$ (Table~\ref{tab:SyntL20}) and $\lambda^{-1}=500$ (Table~\ref{tab:SyntL500}), the highest attainable entropy is about $0.5$ bits below the maximum for $n=2$, but as $n$ increases, the gap increases to about $1$ bit.
In the case for $n=32$ and $\lambda^{-1}=500$, the highest entropy (obtained with $t=15$) is $3.25$, which is $1.75$ bits below the maximum.

Our experiments allow to conclude that, when defining the household groups, we should select their size, $n$, considering the desired level of privacy and assuming that the entropy will be about $1$ bit below the maximum ($\log_2 n$).

The number of time intervals $t$ can not be chosen as it depends on the billing period, but in general $t$ will tend to take high values (one month billing periods are quite usual).

\section{Conclusion}

In this paper an entropy-based measure for quantifying the real privacy provided by anonymous privacy-preserving smart metering methods has been proposed. As the underlying problem required for solving the original problem statement has shown to be too hard, we have performed several experiments on a relaxed formulation of it.

The experiments have shown that the attained privacy, measured as entropy, is about one bit below the theoretical maximum. Privacy increases when all the electricity readings of the involved meters are similar.

\section*{Acknowledgments}

This study was funded by the European Regional Development Fund of the European Union in the scope of the ``Programa Operatiu FEDER de Catalunya 2014--2020'' (project number COMRDI16-1-0060), by the Spanish Ministry of Science, Innovation and Universities (project number MTM2017-83271-R), and by the Federal Ministry for Economic Affairs and Energy of Germany in the SINTEG project DESIGNETZ (project number 03SIN224).

\end{document}